\documentclass[twoside,english,sort&compress]{iopart}
\usepackage{mathptmx}

\usepackage[T1]{fontenc}
\usepackage[latin9]{inputenc}
\usepackage{geometry}
\usepackage{xcolor}
\usepackage{units}
\usepackage{amsbsy}
\usepackage{amstext}
\usepackage{graphicx}
\usepackage[numbers]{natbib}

\makeatletter

\usepackage{iopams}
\usepackage{setstack}

\newcommand{\eqref}[1]{(\ref{#1})}
\makeatother

\usepackage{babel}
\begin{document}

\title[Generalized Fisher-Lee relation]{Mode-resolved transmission functions: an individual Caroli formula}

\author{{\Large{}Hocine Boumrar, Hand Zenia, and Mahdi Hamidi} }

\address{{\large{}Laboratoire de Physique et Chimie Quantique
(LPCQ), Universit\'e Mouloud Mammeri, 15000 Tizi-ouzou, Algeria}}

\ead{{\large{}hand.zenia@ummto.dz}}

\begin{abstract}
\textcolor{black}{Efficient manipulation of energy at the nanoscale
is crucial for advancements in modern computing, energy harvesting,
and thermal management. Specifically, controlling quasiparticle currents
is critical to these ongoing technological revolutions. This work
introduces a novel and physically consistent approach for computing
polarization-resolved transmission functions, a crucial element in
understanding and controlling energy transport across interfaces.
We show that this new method, unlike several previously derived formulations,
consistently yields physically meaningful results by addressing the
origin of unphysical behavior in other methods. We demonstrate that
while multiple decompositions of the transmission function are possible,
only there is a unique and unambiguous route to obtaining physically
meaningful results. We highlight and critique the arbitrary nature
of these alternative decompositions and their associated failures.
While developed within the framework of phonon transport, the individual
Caroli formula is general and applicable to other fermionic and bosonic
quasiparticles, including electrons, and to internal degrees of freedom
such as spin and orbital polarization. Through a comparative analysis
using a simple model system, we validate the accuracy and reliability
of the individual Caroli formula in capturing polarization-specific
transmission properties. This new method provides a more accurate
understanding of both phonon and electron transport, offering novel
ways for optimization of thermoelectric devices and energy-efficient
computing technologies. }
\end{abstract}

\section{Introduction}

Advances in information technology require miniaturization to keep
up with the demand for higher performance. One of the challenges,
however, is efficient dissipation of the increased heat produced as
the result of scaling down the electronic components. Conversely,
in applications of thermoelectricity it is the lack of such efficient
dissipation (or impedance) that is sought after. Indeed a strongly
suppressed thermal conductance is crucial to keep a temperature gradient
across a thermoelectric conversion ``device'', necessary for its
operation\cite{Gang15}. However, the need for both high electrical
conductivity and low thermal conductivity is hard to achieve in conventional
materials. A thorough understanding of how phonons are transmitted
across interfaces is paramount for the design and optimisation of
nanoscale devices.

The problem of thermal transport can be treated at different lengthscales.
On the largest scale the acoustic mismatch model (AMM) and the diffuse
mismatch model(DMM) rely on ``an idealized model of the phonon dispersion''
and ``ignore the contribution from optical phonon modes''\cite{Little59,Swartz89}.
At a ``smaller lengthscale'' the Boltzamnn transport equation (BTE)
is used\cite{Murthy05}. Here two methods are distinguished: gray
and nongray. In the nongray phonon BTE detailed information about
dispersion and polarization of the phonon modes is required to accurately
capture the physics of phonon transport\cite{Zahiri19}. This information
can only be obtained by considering the lowest scale of all. This
is the ``mesoscopic/nanoscopic'' scale where the phonon mean free
path is larger than the ``device'' and ballistic transport is dominent.
At this scale one obtains extra information on the phonon dispersion
and polarization as well as on the dependence of the transmitance
and reflexion on them. One such piece of information is the contribution
of the optic modes to the phonon transport which is completely ignored
at larger lengthscales. The other regards mode-resolved or polarization-resolved
transmittance. The equivalent quantity in optics is the change of
the polarization of light when it passes through a medium. As in optics,
we usually (or experimentalists,users) are not interested in the details
or mechanisms at play in the scattering region. We would like to know
whet happens when a phonon(electron) travelling in a sinlge bulk mode
(Bloch orbital) of one lead is scattered into the modes (orbitals)
that are available in the other lead.

We expect that being able to extract detailed information concerning
the transmission function can have further applications such as in
nanoelectromechanical systems where proteins can not only be identified
by their mass, through shifts in vibration frequency of the modes
of the ``device'' or sensor\cite{Sader24}, but also by their configuration
through the effect on the polarization-resolved transmission or reflection
of the modes.

One can also rely on Molecular Dynamics (MD) to compute phonon transmission
by analyzing the energy flux across an interface or through a structure\cite{Wang08}.
MD simulations can naturally incorporate unharmonicity and work at
high temperatures where classical behaviours dominate\cite{Chen18}.
The simulations can fully include both elastic and inelastic scattering
events, as well as the influence of disorders in the structure\cite{Liao20}.
However, while MD can calculate mode-dependent transmission, it is
challenging to extract detailed, mode-resolved reciprocal-space information
from real-space data\cite{Li19}\cite{Ong20}.

Landauer was the first to find/publish a phenomenological expression
for conductance in a two-terminal setup in terms of transmission probabilities\cite{Landauer57,Landauer70}.
Caroli later on put the formalism on a firm footing by providing a
formula to compute the transmission function given a model system
(in the harmonic approximation for phonons and in the tight-binding
approximation for the electrons)\cite{Caroli71}. The necessary parameters(force
constants for phonons and site energies and hopping and overlap integrals
for the electrons) for the model can nowadays be obtained directly
from first-principles calculations\cite{Klockner18}. At this stage the required
transmission function is obtained using two distinct formalisms: WFM(Wave
Function Matching Mathod)\cite{Ando91,Khomyakov04,Chen19,Boumrar20}
and AGF(Atomistic Green's Function)\cite{Polanco21,Zeng22,Boumrar20}.
While the two yield identical results, it was not clear how they are
related. Recently, however, a rigourous and complete equivalence of
the two methods has been shown\cite{Boumrar20}. In \cite{Boumrar20}
a method of computing polarization-resolved transmission functions
was also given. \textit{ }

Before delving into the details concerning computation of mode-resolved
phonon or orbital-resolved electron transmission, we feel it is necessary
to mention the existence of an alternative way of looking at detailed
information on the transmission function. This is known as the eigenchannel
decomposition\cite{Buttiker88}. It is achieved by diagonalising
a transmission probability matrix, which is derived from the Green's
functions of the system\cite{Klockner18}. The resulting eigenvectors,
or eigenchannels, represent the different pathways that phonons can
take through the device, with corresponding eigenvalues indicating
their transmission probabilities\cite{Klockner18}. The focus is
put, in this method, on the central part on the subspace of which
the scattering states are projected\cite{Klockner18}. Hence, the
difference with the mode-resolved approach where the focus is rather
on the Bloch modes of the leads.

Prior to the method developped in \cite{Boumrar20} several attemps
were made at obtaining mode-resolved transmission functions. The first
was by Huang et al.\cite{Huang11} who computed the polarization
resolved transmittance by diagonalizing the spectral function which
then led to decompsed self-energies. They applied this method to Si/Ge/Si
heterogeneous thin film structure. The second attempt was made by
Ong and Zhang\cite{Ong15} who based their method on an extension
of the AGF formalism. They applied their method to the calculation
of mode-resolved transmittance across the graphene--hexagonal boron
nitride interface ``based on the concept of the Bloch matrix''.
They arrived at a similar expression to one published in \cite{Boumrar20}
and later in .\cite{Jin16}. The third attempt at obtaining mode-resolved
transmitance was developed by Sadasivam et al.\cite{Sadasivam17}
and based on the AGF formalism along with the Lipmann-Schwinger equation
to obtain surface scattering states. They applied their method to
the Si/Ga interface.

In this work we give a rigorous derivation of the polarization-resolved
transmission functions. These, as has already been shown previously
in \cite{Boumrar20} can be obtained from either the formalism of
AGF or that of WFM. In \cite{Boumrar20} the complete equivalence
of the two formalisms has already been demonstrated. The advantage
of the current derivation is two-fold: (a) the steps involved in the
derivation are rigorous and easy to follow, and (b) the amount of
time involved in computing the desired quantities is reduced drastically.
Our analytical demonstrations and the critique of the earlier methods
will be put to the test via a simple model where the comparison is
easy to carry out. We use a square lattice and compute the mode-resolved
transmission functions using the three methods discussed in this work.
The results will be compared to the ones published in \cite{Boumrar20}.

\textcolor{black}{The primary quantity that carries information on
the bulk modes and that appears explicitly in the Caroli forumla for
the transmittance is the escape rate $\Gamma$. Indeed, even though
$\Gamma=i\left(\Sigma-\Sigma^{\dagger}\right)$ is written in terms
of the surface self-energy $\Sigma$, the latter appears only implicitly,
through the device Green's function $G_{C}$, in the Caroli formula.
As such, any attempt at resolving the transmittance in terms of the
bulk modes must necessarily lead to decomposed escape rates. There
are many seemingly correct ways the escape rate can be decomposed.
It appears, however, and for reasons that will be exposed later/below
that only one such method leads to physically meaningful results.
The others, although mathematically sound ``in some cases'', yield
results that are clearly unphysical, such as negative transmission
amplitudes, or transmittances that are greater than unity. In the
following we start with a survey of earlier methods available in the
literature and comment on their shortcomings, some physical and some
even mathematical in nature. We then expose our method and argue for
its validity and simplicity from the mathematical standpoint. We also
make the point why it leads to physically sound and meaningful results.
In order to further illustrate the discussion and comparison of the
different methods, we consider the simplest case where the merits
of each method can be brought out. The goal is to allow for other
workers to reproduced our results without the further complications
that may arise from using realistic models where the numerics or the
parametrisation of the interactions may come in the way of focusing
on the main issues at hand.}

The paper is organized as follows. In the next section we provide
a description of a general setup of a two-terminal device, and an
overview of the quantities that are central to understand the derivations
put forth by the methods. These will be discussed/reviewed in the
third section; where our method is also presented. In the fourth section
the methods are applied to a simple model for a numerical calculation
of the mode-resolved transmittance functions; where the results are
used as a means to validate our comparison of the methods. The main
arguments and conclusions are presented in the last section. 

\section{The Setup}

\begin{figure}
\begin{centering}
\includegraphics[scale=0.4]{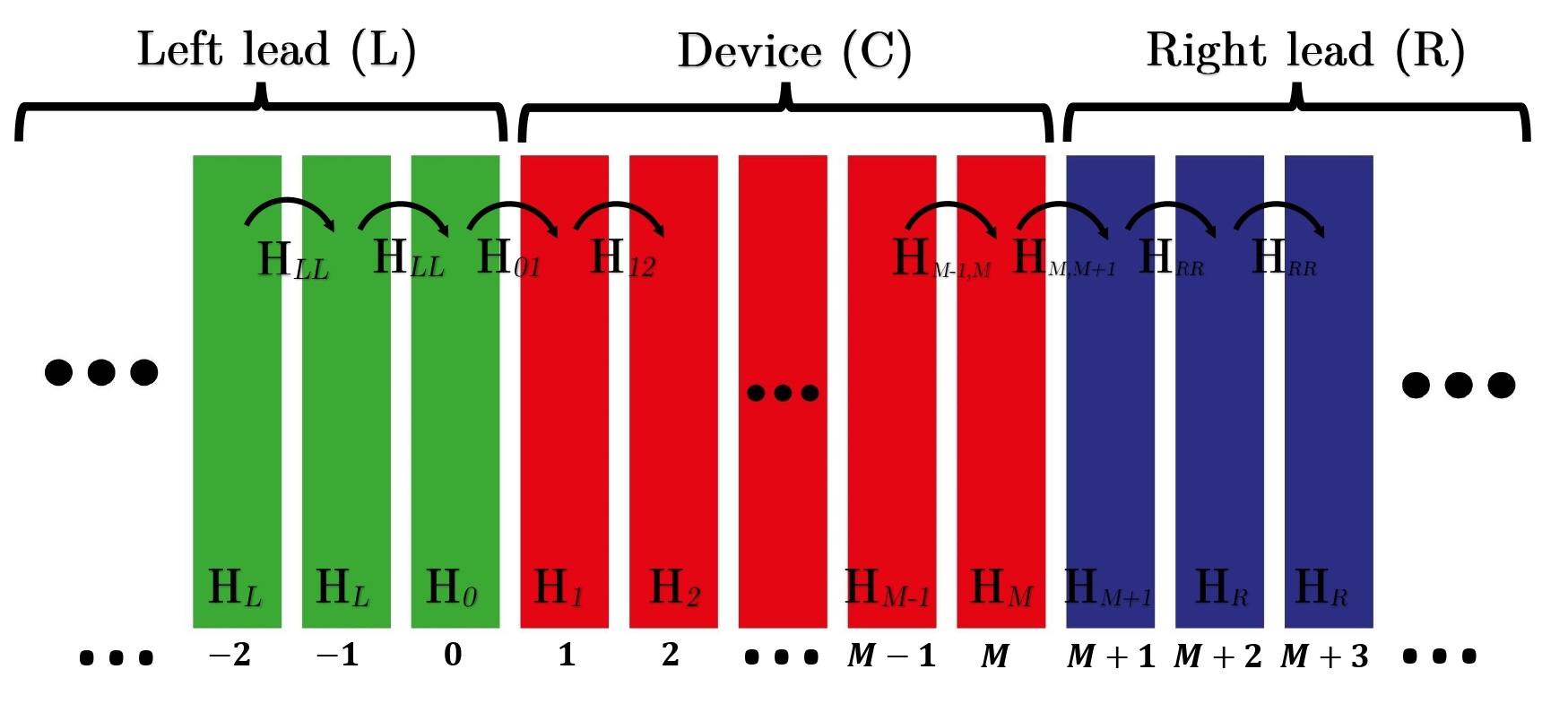}
\par\end{centering}
\begin{centering}
\caption{Schematic representation of the two-terminal system. It is divided
into slices whose corresponding Hamiltonians are also shown.\label{fig:setup}}
\par\end{centering}
\end{figure}

The total Hamiltonian $\mathbf{H}$ of the ``device'', as shown
in Fig. \eqref{fig:setup}, and left and right leads is given by a
tridiagonal ``supermatrix'':
\begin{equation}
\mathbf{H}=\left[\begin{array}{ccccccc}
\ddots & \ddots & 0 & 0 & 0 & 0 & 0\\
\ddots & {\bf {H}_{\textit{L}}} & {\bf {H_{\textit{LL}}}} & 0 & 0 & 0 & 0\\
0 & {\bf {H_{\textit{LL}}^{\dagger}}} & {\bf {H_{\textit{L}}}} & {\bf {H_{\textit{LL}}}} & 0 & 0 & 0\\
0 & 0 & {\bf {H_{\textit{LL}}^{\dagger}}} & {\bf {H_{\textit{C}}}} & {\bf {H_{\textit{RR}}}} & 0 & 0\\
0 & 0 & 0 & {\bf {H_{\textit{RR}}^{\dagger}}} & {\bf {H_{\textit{R}}}} & {\bf {H_{\textit{RR}}}} & 0\\
0 & 0 & 0 & 0 & {\bf {H_{\textit{RR}}^{\dagger}}} & {\bf {H_{\textit{R}}}} & \ddots\\
0 & 0 & 0 & 0 & 0 & \ddots & \ddots
\end{array}\right],\label{totalham}
\end{equation}
where ${\bf H_{\textit{L}}}$ and ${\bf {H}_{\textit{R}}}$ are, respectively,
the Hamiltonan of a single slice in the left and right leads;${\bf {H_{\textit{LL}}}}$
and ${\bf {H_{\textit{RR}}}}$ are the ``force constants'' between
two neighboring slices in the left and right leads, respectively;
and ${\bf {H_{\textit{C}}}}$ is the Hamiltonian of the central part
attached to the leads. Note that each of these quantities are in general
$n\times n$ matrices, where $n$ is the number of the degrees of
freedom and also of the normal modes.

First, the leads are treated as semi-infinite systems and their respective
surface GFs $g_{L}$ and $g_{R}$ are computed, mostly using a well-known
algorithm due to Lopez-Sancho{[}REFS{]}. Then, the effect of the presence
of the leads on the ``central region'' is accounted for by left
and right self-energies defined as

\begin{equation}
\boldsymbol{\Sigma}_{\textit{L}}={\bf {H_{\textit{LL}}^{\dagger}}}{\bf {g}_{\textit{L}}}{\bf {H_{\textit{LL}}}}\label{sigmaL}
\end{equation}
and
\begin{equation}
\Sigma_{\textit{R}}={\bf {H_{\textit{RR}}}}{\bf {g}_{\textit{R}}}{\bf {H_{\textit{RR}}^{\dagger}}}.\label{sigmaR}
\end{equation}
These are now added to the first and last diagonal elements of $H_{C}$
to give $H_{C}^{eff}$. As a result $H_{C}^{eff}$ is no longer Hermitian,
a sign that the central region is to be treated as an open system.
Inversion of $H_{C}^{eff}$ then yields $G_{C}$. Two further important
quantities are defined 
\begin{equation}
\boldsymbol{\Gamma}_{\textit{L(R)}}=i(\boldsymbol{\Sigma}_{\textit{L(R)}}-\boldsymbol{\Sigma}_{\textit{L(R)}}^{\dagger}),\label{eq:escape_rates}
\end{equation}
and are called ``escape rates''. Given all the quantities, the total
transmission function is given by the Caroli formula
\begin{equation}
t=\mathrm{Tr}\left(\boldsymbol{\Gamma}_{\textit{L}}{\bf {G}_{\textit{C}}\boldsymbol{\Gamma}_{\textit{R}}{\bf {G}_{\textit{C}}^{\dagger}}}\right).\label{eq:caroli_form}
\end{equation}

All the methods discussed below aim at obtaining mode-resolved escape
rates $\boldsymbol{\Gamma}_{\textit{L,i}}$ and $\boldsymbol{\Gamma}_{\textit{R,i}}$.
These are then plugged into the above Caroli formula to obtain mode-resolved
transmittance functions:

\begin{equation}
{\textit{t}}_{i,j}=\mathrm{Tr}\left(\boldsymbol{\Gamma}_{\textit{L,i}}{\bf {G}_{\textit{C}}\boldsymbol{\Gamma}_{\textit{R,j}}{\bf {G}_{\textit{C}}^{\dagger}}}\right)\label{eq:mode_res_t_caroli}
\end{equation}

\section{Earlier Derivations}

\textcolor{black}{As mentioned earlier, any attempt at obtaining mode-resolved
transmitance functions ought to lead to resolved escape rates. In
this section we review two methods that have been published in the
literature \cite{Huang11,Sadasivam17}, present a method to correct
one of these, and elaborate on a third method devolloped by us to
further illustrate the reason behind the failures of the earlier ones,
even when the decomposition is seemingly correctly carried out. The
first methods begin by decomposing the spectral function\cite{Huang11},
the second starts from the decomposition of the surface Green's functions\cite{Sadasivam17}.
Given that the latter is incomplete, we show how to correct it following
the original line of reasoning of its authors. We then exposed a third
possible method that consists of starting the decomposition from quantities
known as the Bloch matrices.}

\subsection{Method I: Decomposing the Spectral Function}

In this method due to Huang et al.\cite{Huang11} the mode-resolved
escape rates are obtained simply by diagonalizing the spectral function
$A$ obtained from the surface GFs $g_{L}$ and $g_{R}$. It starts
with theses expressions of the escape rates :
\begin{equation}
\left\{\begin{array}{lr}
\boldsymbol{\Gamma}_{\textit{L}}= & {\bf {H_{\textit{LL}}^{\dagger}}{\textit{\textbf{A}}}_{\textit{L}}{\bf {H_{\textit{LL}}}}},\\
\boldsymbol{\Gamma}_{\textit{R}}= & {\bf {H_{\textit{RR}}}{\textit{\textbf{A}}}_{\textit{R}}{\bf {H_{\textit{RR}}^{\dagger}}}}.
\end{array}\right.
\label{taux de transfert2}
\end{equation}
where $H_{LL}$ and $H_{RR}$ are the left and right matrices connecting
two adjacent slices of the leads. The matrix ${\textit{\textbf{A}}}_{\textit{L(R)}}$
left or right spectral function is defined as 
\begin{equation}
{\textit{\textbf{A}}}_{\textit{L(R)}}=i({\bf {g}_{\textit{l(r)}}}-{\bf {g}_{\textit{l(r)}}^{\dagger}}),\label{fonction spectral}
\end{equation}
with ${\bf {g}_{\textit{l(r)}}}$ being the left and right surface
Green's functions. The spectral function is then diagonalized
\begin{equation}
{\textit{\textbf{A}}}=\sum_{\textit{i=1}}^{n}\lambda_{i}\boldsymbol{\phi}_{i}\boldsymbol{\phi}_{i}^{\dagger},\label{fonction spectral2}
\end{equation}
where $\boldsymbol{\phi}$ and $\lambda$ are eigenvectors and eigenvalues
and $n$ is the number of normal modes. This ``decomposed'' matrix
is plugged into the above equation, from which the individual escape
rates are obtained :
\begin{equation}
\left\{\begin{array}{lr}
\boldsymbol{\Gamma}_{\textit{L,i}}= & {\bf {H_{\textit{LL}}^{\dagger}}\lambda_{L,i}\boldsymbol{\phi}_{L,i}\boldsymbol{\phi}_{L,i}^{\dagger}{\bf {H_{\textit{LL}}}}},\\
\boldsymbol{\Gamma}_{\textit{R,i}}= & {\bf {H_{\textit{RR}}}\lambda_{R,i}\boldsymbol{\phi}_{R,i}\boldsymbol{\phi}_{R,i}^{\dagger}{\bf {H_{\textit{RR}}^{\dagger}}}}.
\end{array}\right.
\label{taux de transfert3}
\end{equation}
where $\boldsymbol{\Gamma}_{\textit{L,i}}$ and $\boldsymbol{\Gamma}_{\textit{R,i}}$
are the individual escape rates that correspond to the left and right
leads. These quantities are subsequently plugged into the Caroli formula
(Eq. \ref{eq:mode_res_t_caroli}) to obtain the mode-resolved transmittance
$t_{i,j}$ between mode $i$ on the left and mode $j$ on the right.

At this point, it is important to point out that apparently there
is no connection between the eigenvalue $\lambda_{i}$ obtained here
and the $i^{\mathrm{th}}$ mode in the lead. The quantities $t_{i,j}$
therefore cannot represent mode-resolved transmittance between mode
$i$ on the left and mode $j$ on the right as claimed in \cite{Huang11}.

\subsection{Method II: Decomposing the Surface Green's Functions}

\subsubsection{Original Formulation}

The method developed by Sadasivam \textit{et al.}\cite{Sadasivam17}
begins by decomposing the surface Green's functions into mode-resolved
ones starting from ``surface modes'' $\psi_{L}$ and $\psi_{R}$.
These are obtained from the original bulk modes $\boldsymbol{\varphi}_{L/R}$
using the Lippmann-Schwinger equation.  They use a different method
to compute the surface Green's function: a perturbative approach of
the Dyson equation, where the unperturbed Hamiltonian $H_{0}$ is
that of the perfect lead, taken as an infinite bulk; and the perturbation
$V$ is such that it breaks/suppresses interactions between the $n^{\mathrm{th}}$
and $n+1^{\mathrm{st}}$ layer of the bulk, where then they become
surface layers. This means that $V_{n,n+1}=-H_{LL}$ for the left
lead and $V_{n,n+1}=-H_{RR}$ for the right lead, with all other $V_{kl}$
terms vanishing. The surface is represented by the $n^{\mathrm{th}}$
layer for the left lead, and by the $n+1^{\mathrm{st}}$ layer for
the right lead. In both cases the perturbation is non-zero only between
the layers $n$ and $n+1$. Through the Lipmann-Schwinger equation,
they obtain the surface ``modes'' for the left lead as
\[
\boldsymbol{\psi}_{\textit{L}}=\boldsymbol{\boldsymbol{\varphi}}_{\textit{n}}-\textbf{\textit{g}}_{\textit{L}}\boldsymbol{H_{\textit{LL}}}\boldsymbol{\boldsymbol{\varphi}}_{\textit{n+1}},
\]
and for the right lead as
\[
\boldsymbol{\psi}_{\textit{R}}=\boldsymbol{\boldsymbol{\varphi}}_{\textit{n+1}}-\textbf{\textit{g}}_{\textit{R}}\boldsymbol{H}_{RR}^{\dagger}\boldsymbol{\boldsymbol{\varphi}}_{\textit{n}}.
\]
The surface Green's functions are decomposed as
\begin{equation}
\left\{\begin{array}{lr}
\boldsymbol{g}_{L,i}(\omega;k_{y})= & -\frac{ia}{2}\frac{\boldsymbol{\psi}_{\textit{L,i}}\boldsymbol{\psi}_{\textit{L,i}}^{\dagger}}{2\omega\upsilon_{L,i}}\\
\boldsymbol{g}_{R,j}(\omega;k_{y})= & \frac{ia}{2}\frac{\boldsymbol{\psi}_{\textit{R,\ensuremath{j}}}\boldsymbol{\psi}_{\textit{R,\ensuremath{j}}}^{\dagger}}{2\omega\upsilon_{R,j}}
\end{array}\right.
\label{eq:sad_res_surf_gfs}
\end{equation}
where the indices $i$ and $j$ enumerate the normal modes in the
left and right leads, respectively, and $v_{L/R}$ are generalized
group velocities\cite{Allen79,Chang82}.  These decomposed surface
GFs are to obtain decomposed self-energies as in eqs. \ref{sigmaL}
and \ref{sigmaR}, from which the decomposed escape rates are deduced
(eq. \ref{eq:escape_rates}). Then eq. \ref{eq:mode_res_t_caroli}
is used to compute the polarization-resolved transmittance. Again,
once these quantities are plugged into the Caroli formula as in Eq.
\ref{eq:mode_res_t_caroli} to obtain $t_{i,j}$ the mode-resolved
transmittance between mode $i$ on the left and mode $j$ on the right.

The claim here is that since only propagating eigenvectors enter the
definition of the surface modes, only the corresponding propagating
bulk modes contribute to the transmittance. This is questionable since
the surface GFs, and the bulk ones incidentally, that enter in $\psi_{L/R}$
must contain information on both propagating and evanescent modes.
Moreover, the surface $\psi_{L/R}$ is in fact a mixture of bulk modes,
and as such there is no one-to-one correspondence between surface
and bulk modes.

\textcolor{black}{We carried out numerical calculations using the
mode-resolved surface GFs as given in Sadasivam; and we obtained completely
incorrect results for the mode-resolved transmission function $t_{i,j}$.
Even the total transmission, when computed as the sum of the mode-resolved
ones, turned out to be incorrect. Our explanation is that the ``total'',
and correctly computed, surface GF $g_{L}$ is not identical to the
sum of the mode-resolved functions $g_{L/R,i}$ as given in eq. \ref{eq:sad_res_surf_gfs}.
For this reason the method as originally formulated in \cite{Sadasivam17}
has not been used in the numerical application below. Instead, we
obtained a corrected formulation along the lines of ref. \cite{Sadasivam17}
for the comparison. The corrected formula is discussed next.}

\subsubsection{Corrected Formulation}

We have pointed earlier that the formulation put forth in ref. \cite{Sadasivam17}
was flawed and here we provide our correction. Indeed, we obtained,
in the spirit of Sadasivam et al.'s work\cite{Sadasivam17}, the
correct formula for the mode-resolved surface GFs. Using their notation,
and Dyson's equation, the surface Green's function can be written
as
\[
g_{L}=G_{n,n}-g_{L}H_{LL}G_{n+1,n},
\]
with
\[
G_{n+1,n}=B_{R}G_{n,n},
\]
where, obviously, the last quantities are defined for the perfect
leads. Here $B_{L/R}$ are the Bloch matrices defined in eq. \ref{eq:bloch_matrices}.
Given, on the other hand, that the unperturbed ``surface'' Green's
function can be written, in terms of the bulk modes, as 
\[
G_{n,n}=-i\sum_{i}\frac{\boldsymbol{\varphi}_{L,i}\left(\boldsymbol{\varphi}_{L,i}^{a}\right)^{\dagger}}{v_{L,i}},
\]
where the index $i$ enumerates the bulk modes and $v_{L,i}$ is the
corresponding ``generalized'' group velocity, we arrive at
\begin{equation}
g_{L}= \left[I-g_{L}H_{LL}B_{L}\right]G_{n,n} =-i\sum_{i} \left[I-g_{L}H_{LL}B_{L}\right]\frac{\boldsymbol{\varphi}_{L,i}\left(\boldsymbol{\varphi}_{L,i}^{a}\right)^{\dagger}}{v_{L,i}} =  \sum_{i}g_{L,i}.
\end{equation}
Thus, we obtain, from identification,
\[
g_{L,i}=-i\left[I-g_{L}H_{LL}B_{L}\right]\boldsymbol{\varphi}_{L,i}\frac{(\boldsymbol{\varphi}_{L,i}^{a})^{\dagger}}{v_{L,i}}.
\]
While the prefactor on the right-hand side is identical to Sadasivam
et al's\cite{Sadasivam17} $\psi_{n}$, the fraction differs from
their $\psi_{n}^{*}$. The other important difference is the extra
$1/2$ term in their eq. 14. Seemingly, using the corrected formula
one expects improved, if not correct, results. As we will see in the
numerical application below, however, this is not the case, again
due to the presence of the surface $g_{L/R}$ in the mode resolved
$g_{L/R,i}$ itself. Indeed, as mentioned above the surface GF $g_{L/R}$
mixes the bulk modes and also contains information on the evanescent
modes. As a result $g_{L/R,i}$ does not correspond solely to the
$i^{th}$ bulk mode alone. One possible route is to start with mode-resolved
Bloch matrices to obtain decomposed surface GFs $g_{L/R}$. This method
is discussed in the next sub-section.

\subsection{Method III: Decomposing the Bloch matrices}

Starting from the Bloch matrices (see Appendix in \cite{Boumrar20})
\begin{equation}
\begin{array}{lr}
{\bf {B}_{\textit{L}}}= & \textbf{C}_{\textit{L}}~\boldsymbol{\Lambda}_{\textit{L}}~{\bf {C}_{\textit{L}}}^{-1}\\
{\bf {B}_{\textit{R}}}= & \textbf{C}_{\textit{R}}~\boldsymbol{\Lambda}_{\textit{R}}~{\bf {C}_{\textit{R}}}^{-1},
\end{array}
\label{eq:bloch_matrices}
\end{equation}
where $\textbf{C}_{\textit{L(R)}}$ and $\boldsymbol{\Lambda}_{\textit{L(R)}}$
are the left(L) and right(R) eigenvectors and eigenvalues, we compute
the surface GFs using 
\begin{equation}
g_{L}=B_{L}\,H_{LL}^{-1},
\end{equation}
for the left surface GF, for instance, with a similar expression for
the right one. We can decompose the Bloch functions by using individual
eigenvectors and eigenvalues, instead of the corresponding matrices
\begin{equation}
\begin{array}{lr}
{\bf {B}_{\textit{L}}}= & \sum_{i}{\bf {B}_{\textit{L,i}}}=\sum_{i}\textbf{\ensuremath{\boldsymbol{\varphi}}}_{\textit{L,i}}~{\lambda}_{\textit{L,n}}~{\bf {\boldsymbol{\varphi}}_{\textit{L,i}}}^{-1}\\
{\bf {B}_{\textit{R}}}= & \sum_{i}{\bf {B}_{\textit{R,i}}}=\sum_{i}\textbf{\ensuremath{\boldsymbol{\varphi}}}_{\textit{R,i}}~{\lambda}_{\textit{R,i}}~{\bf {\boldsymbol{\varphi}}_{\textit{R,i}}}^{-1},
\end{array}
\label{resolved_bloch_matrices}
\end{equation}
and obtain the mode-resolved GFs. 
\begin{equation}
\begin{array}{lr}
g_{L,i}= & \textbf{\ensuremath{\boldsymbol{\varphi}}}_{\textit{L,i}}~{\lambda}_{\textit{L,i}}~{\bf {\boldsymbol{\varphi}}_{\textit{L,i}}}^{-1}\,H_{LL}^{-1},\\
g_{R,i}= & \textbf{\ensuremath{\boldsymbol{\varphi}}}_{\textit{R,i}}~{\lambda}_{\textit{R,i}}~{\bf {\boldsymbol{\varphi}}_{\textit{R,i}}}^{-1}\,(H_{RR}^{\dagger})^{-1}
\end{array}
\end{equation}
These expressions are to be compared with those obtained by Sadasivam
\textit{et al.}\cite{Sadasivam17}. The individual self-energies
are also obtained from the Bloch matrices as 
\begin{equation}
\begin{array}{lr}
\boldsymbol{\Sigma}_{\textit{L,i}}= & {\bf {H}_{\textit{RR}}~{\bf {B}_{\textit{R,i}}}}\\
\boldsymbol{\Sigma}_{\textit{R,i}}= & {\bf {H}_{\textit{LL}}^{\dagger}~{\bf {B}_{\textit{L,i}},}}
\end{array}
\label{individual_sigmas}
\end{equation}
from which the mode-resolved escape rates are calculated as 
\begin{equation}
\begin{array}{lr}
\boldsymbol{\Gamma}_{\textit{L,i}}= & i(\boldsymbol{\Sigma}_{\textit{L,i}}-\boldsymbol{\Sigma}_{\textit{L,i}}^{\dagger})\\
\boldsymbol{\Gamma}_{\textit{R,i}}= & i(\boldsymbol{\Sigma}_{\textit{R,i}}-\boldsymbol{\Sigma}_{\textit{R,i}}^{\dagger}).
\end{array}
\end{equation}
The final formulae for these are 
\begin{equation}
\begin{array}{lr}
\boldsymbol{\Gamma}_{\textit{R,i}}= & i(H_{R,R}\textbf{\ensuremath{\boldsymbol{\varphi}}}_{\textit{R,i}}{\lambda}_{\textit{R,i}}~{\bf {\boldsymbol{\varphi}}_{\textit{R,i}}}^{-1}-c.c.)
\end{array}
\label{eq:bloch_ind_esc_rates}
\end{equation}
When inserted into the Caroli formula, these are expected to yield
the mode-resolved transmission functions $t_{i,j}$ as in Eq. \ref{eq:mode_res_t_caroli}.
However, as we will see below, this formula is not entirely correct.
Our numerical calculations show that the mode-resolved transmission
functions obtained from the above formula are plagued with similar
problems encountered when using Sadasivam \textit{et al}.'s\cite{Sadasivam17}
original formulation. In the next section we treat a different route
which consists of obtaining mode-resolved escape rates directly from
individual group velocities. It is the main contribution of this work.

\section{Method IV: Decomposing the Escape Rates}

In this section we proceed to the derivation of the correct method
leading to results in agreement with those obtained in earlier literature\cite{Boumrar20,Ong15}.
As shown above, starting from the decomposition of the Bloch matrices
we obtain mode-resolved escape rates that when plugged into the Caroli
formula lead to what seem a legitimate polarisation-resolved transmitivities.
When implemented numerically, however, the results are unphysical
in the same way as seen with the methods Huang et al.\cite{Huang11}
and of Sadasivam \textit{et al.}\cite{Sadasivam17}. Besides, and
as we will see now it turns out that the decomposed escape rates when
carried out ``properly'' contain extra terms that are absent in
Eq. \ref{eq:bloch_ind_esc_rates}. It is this key difference does
not show up in one dimension or when the system is set up such that
it behaves as if it were one-dimensional.

Our method here consists in computing the mode-resolved escape rates
starting from group velocities as follows. We begin by writing the
escape rates in terms of the group velocities and the generalized
eigenvectors as
\[
\begin{array}{ccc}
\boldsymbol{\Gamma}_{\textit{L}} & = & (\textbf{C}_{\textit{L}}^{a\dagger})^{-1}\textbf{V}_{\textit{L}}^{a}(\textbf{C}_{\textit{L}}^{a})^{-1}\\
\boldsymbol{\Gamma}_{\textit{R}} & = & (\textbf{C}_{\textit{R}}^{\dagger})^{-1}\textbf{V}_{\textit{R}}(\textbf{C}_{\textit{R}})^{-1},
\end{array}
\]
as derived in \cite{Boumrar20}. Given the group velocity matrices
are diagonal, we can decompose the escape rates as
\[
\begin{array}{ccc}
\boldsymbol{\Gamma}_{\textit{L}} & = & \sum_{i}{\bf {\bf \boldsymbol{\Gamma}}_{\textit{L,i}}}\\
\boldsymbol{\Gamma}_{\textit{R}} & = & \sum_{i}{\bf {\bf \boldsymbol{\Gamma}}_{\textit{R,i}}},
\end{array}
\]
with
\[
\begin{array}{ccc}
\boldsymbol{\Gamma}_{\textit{L,i}} & = & (\boldsymbol{\varphi}_{\textit{L,i}}^{a\dagger})^{-1}{\bf {V}_{\textit{L}}^{\textit{a}}\textit{(i,i)}({\boldsymbol{\boldsymbol{\varphi}}}_{\textit{L,i}}^{\textit{a}})^{-1}}\\
\boldsymbol{\Gamma}_{\textit{R,i}} & = & (\boldsymbol{\varphi}_{\textit{R,i}}^{\dagger})^{-1}{\bf {V}_{\textit{R}}\textit{(i,i)}(\boldsymbol{\varphi}_{\textit{R,i}})^{-1},}
\end{array}
\]
where the inverse of an eigenvector is actually its pseudoinverse.

With the expressions derived in\cite{Boumrar20} for the group velocities,
\[
\begin{array}{ccc}
v_{L}^{a}(i,j) & = & i\delta_{i,j}\left[\left(\mathbf{\boldsymbol{\varphi}}_{L,i}^{a}\right)^{\dagger}H_{LL}\boldsymbol{\varphi}_{L,i}^{a}\tilde{\lambda}_{L,i}^{a}-\left(\mathbf{\boldsymbol{\varphi}}_{L,i}^{a}\right)^{\dagger}H_{LL}^{\dagger}\mathbf{\boldsymbol{\varphi}}_{L,i}^{a}\left(\tilde{\lambda}_{L,i}^{a}\right)^{\dagger}\right]\\
v_{R}(i,j) & = & i\delta_{i,j}\left[\left(\mathbf{\boldsymbol{\varphi}}_{R,i}\right)^{\dagger}H_{RR}\mathbf{\boldsymbol{\varphi}}_{R,i}\lambda_{R,i}-\left(\mathbf{\boldsymbol{\varphi}}_{R,i}\right)^{\dagger}H_{RR}^{\dagger}\mathbf{\boldsymbol{\varphi}}_{R,i}\left(\lambda_{R}\right)^{\dagger}\right]
\end{array}
\]
we arrive at the polarisation resolved escape rates 
\begin{equation}
\begin{array}{ccc}
\boldsymbol{\Gamma}_{\textit{L,i}} & = & i\left(\left(\boldsymbol{\varphi}_{L,i}^{a}\right)^{\dagger}\right){}^{-1}\left[\left(\boldsymbol{\varphi}_{L,i}^{a}\right)^{\dagger}H_{LL}\boldsymbol{\varphi}_{L,i}^{a}\tilde{\lambda}_{L,i}^{a}-c.c.\right]\left(\boldsymbol{\varphi}_{L,i}^{a}\right){}^{-1}\\
\boldsymbol{\Gamma}_{\textit{R,i}}
 & = & i(\boldsymbol{\varphi}_{R,i}^{\dagger})^{-1}\left[\boldsymbol{\varphi}_{R,i}^{\dagger}H_{R,R}\boldsymbol{\varphi}_{\textit{R,i}}\lambda_{\textit{R,i}}-c.c.\right]\boldsymbol{\varphi}_{R,i}^{-1}
\end{array}\label{eq:resolved_esc_rates_gr_vel}
\end{equation}

Again as with the previous methods these are plugged into Eq. \eqref{eq:mode_res_t_caroli}
to obtain mode-resolved transmittances. We not that the previous formula
has extra terms when compared to Eq. \eqref{eq:bloch_ind_esc_rates}.
And it turns out that these are needed to obtain the correct mode-resolved
transmission functions. These are computed next in the numerical application
and checked against the other methods and the results obtained in
\cite{Boumrar20}.

Comparison of eq. \ref{eq:resolved_esc_rates_gr_vel} and eq. \ref{eq:bloch_ind_esc_rates}
shows that they differ by an extra term of the form $\mathbf{\boldsymbol{\varphi}}_{i}^{-1}\mathbf{\boldsymbol{\varphi}}_{i}$
where $\mathbf{\boldsymbol{\varphi}}_{i}^{-1}$ is the pseudo-inverse
of $\mathbf{\boldsymbol{\varphi}}_{i}$. It turns out that this term
is crucial when the eigenvectors $\mathbf{\boldsymbol{\varphi}}_{i}$
are not mutually orthogonal. This happens when the displacement cannot
be resolved into longitudinal and transverse along the direction of
the propagation of the ``current''. We think that this decomposition
is the only sensible one because the information needed to describe
the phonon current must contain both the displacements and the corresponding
group velocity. In the other methods, only partial information about
the modes is retained.

\section{Numerical application}

\subsection{Model}

\begin{figure}[th]
\begin{centering}
\includegraphics[scale=0.25]{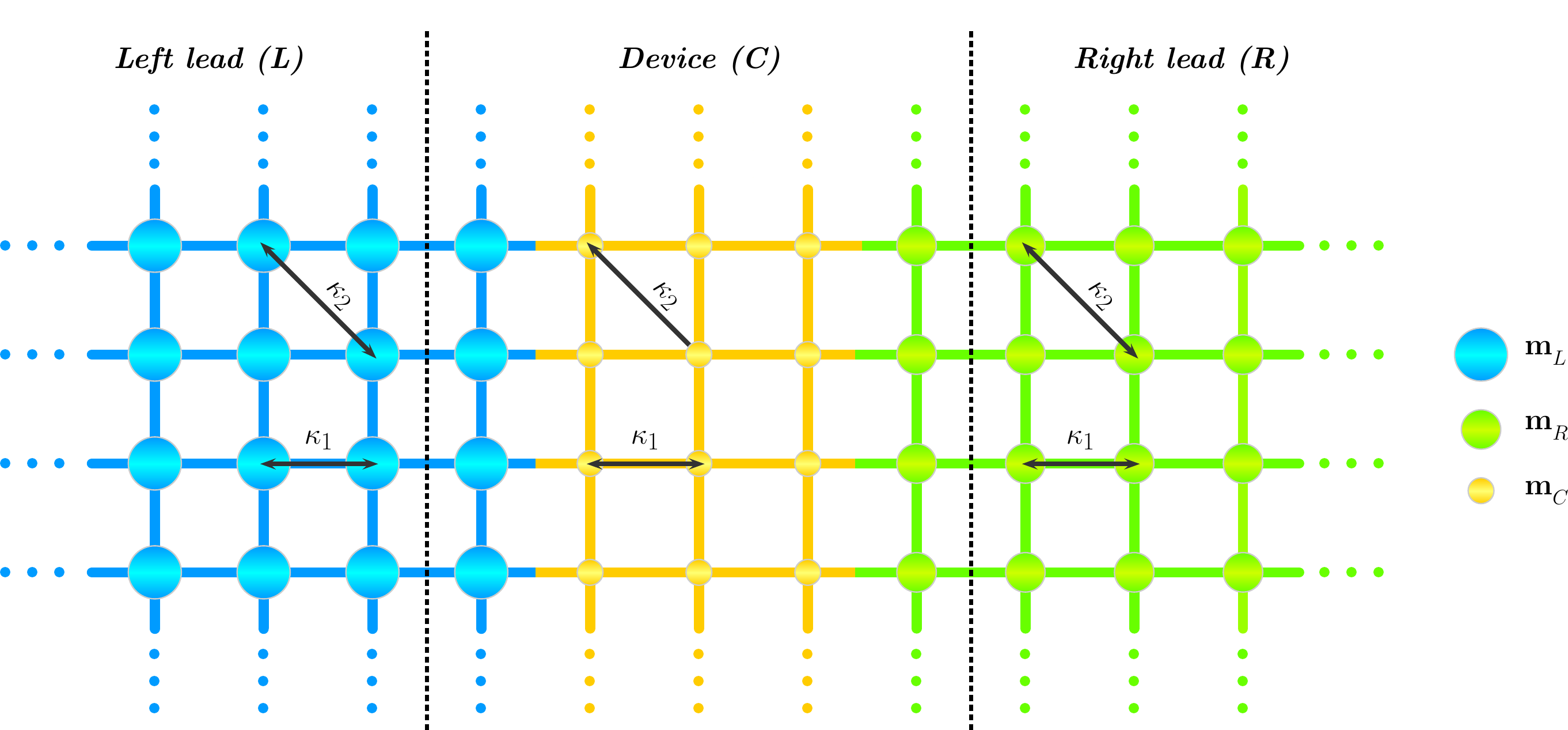}
\par\end{centering}
\caption{A schematic representation the system consisting of a square lattice
with the same lattice constant and force constants ($\kappa_{1}$
for first-neighbor and $\kappa_{2}$ for second-neighbor) throughout.
The masses are different in the three regions: they are $m_{L}$ in
the left lead, $m_{C}$ in the central region, and $m_{R}$ in the
right lead\label{fig:square_lattice_model}.}
\end{figure}

In this section we compare different methods of obtaining the desired
polarization-resolved transmission functions. Most methods as mentioned
earlier rely on resolving one quantity or the other. In some cases
it is easy to show when the method fails, but in others we can only
rely on numerical simulations to carry out this task. The standard
to which all the methods, discussed in this work, are compared is
the method published earlier in\cite{Boumrar20}. As in ref. \cite{Boumrar20}
the merits of each method is gauged by illustrating its application
with a simple model, and also by comparing it to previous common methods.
We consider the case of the 2D square lattice, of lattice constant
$a$, with the central part made up of three atomic planes as shown
in Fig \ref{fig:square_lattice_model}. The force constants are identical
throughout the system, but we choose to use a different mass for each
part of the device. The force constants are $\kappa_{1}$ and $\kappa_{2}=0.5\;\kappa_{1}$
for the first and second-neighbor interactions, respectively. The
masses are $m_{L}$ for the left lead, $m_{R}=0.6\;m_{L}$ for the
right lead, and $m_{C}=0.8\;m_{L}$ for the central part. We define
the frequency unit $\omega_{0}=\sqrt{\kappa_{1}/m_{L}}$. Henceforth
the frequency $\omega$ is given in units of $\omega_{0}$.

\subsection{Results}

\begin{figure}[h]
\begin{centering}
\includegraphics[scale=0.5]{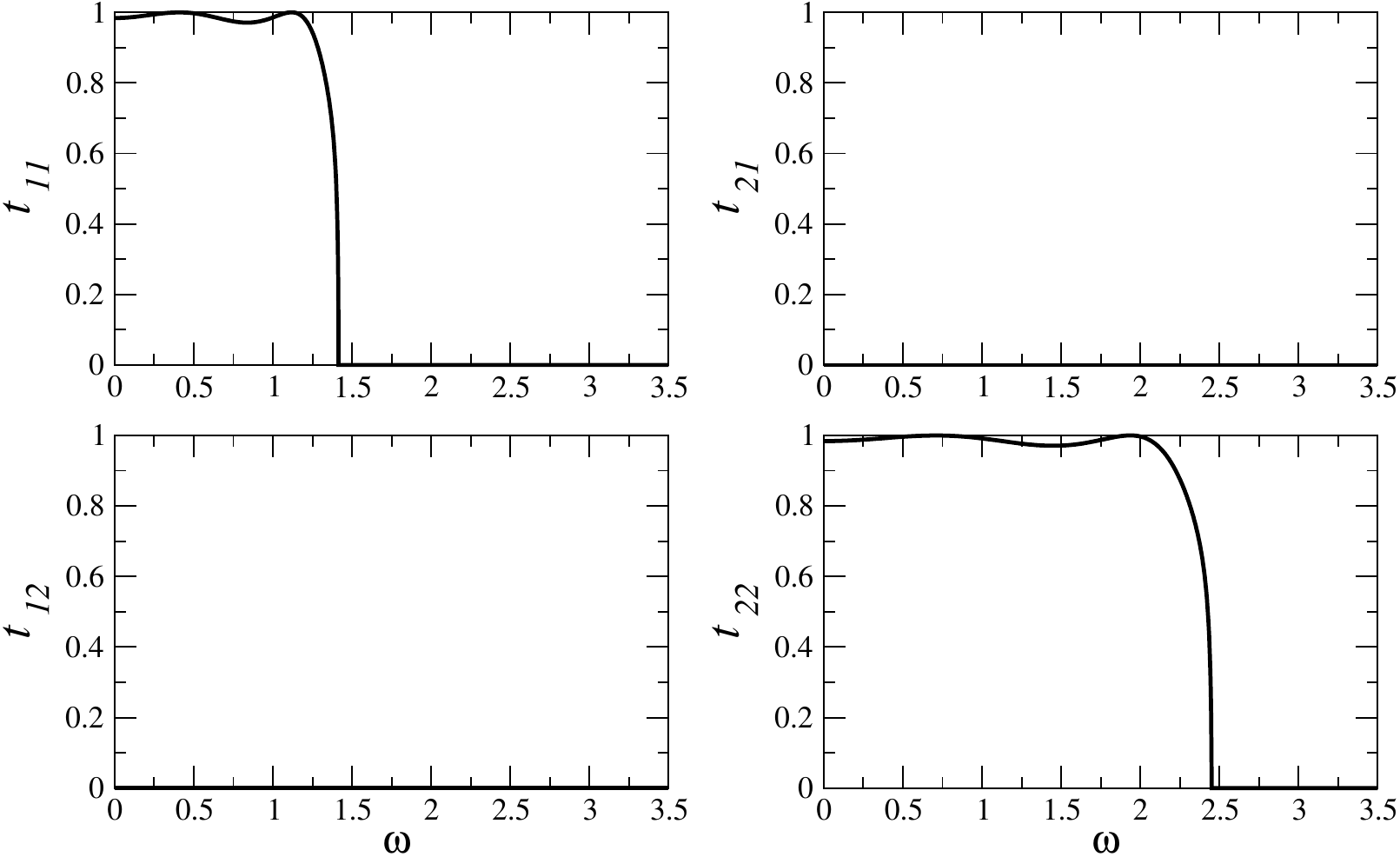}
\par\end{centering}
\caption{Polarization-resolved transmission functions obtained for $k_{y}=0$.\label{fig:results_all_ky_0}}
\end{figure}

\begin{figure}[h]
\begin{centering}
\includegraphics[scale=0.5]{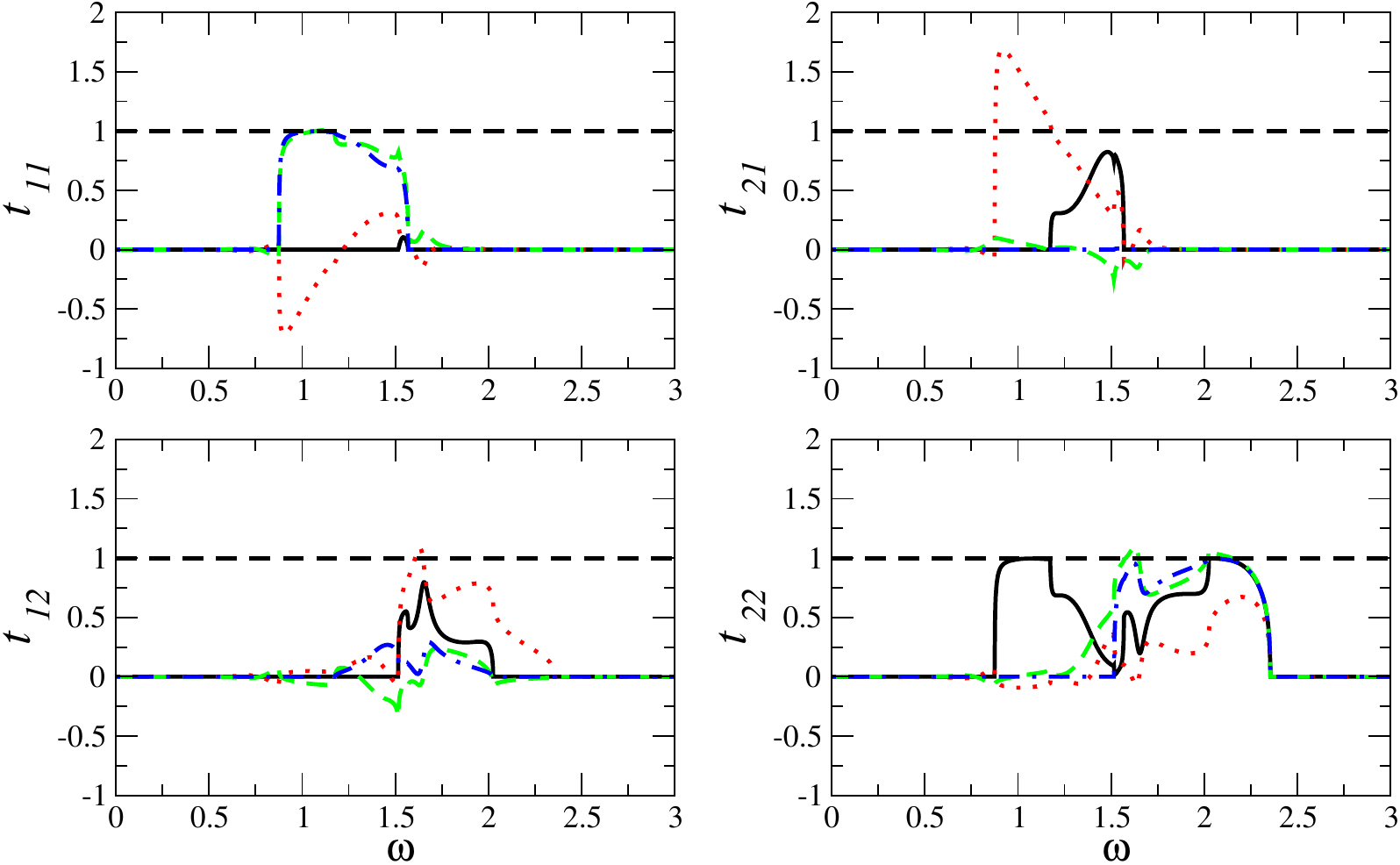}
\par\end{centering}
\caption{Polarization-resolved transmission functions obtained using Huang's
method(black continuous), corrected Sadasivam's(red dotted), Bloch
matrix decomposition(green dashed), and escape rates decomposition(blue
dot-dashed)., for $k_{y}=1$.\label{fig:results_all_ky_1}}
\end{figure}

Results of our calculations using the methods outlined in this work
are displayed in Figs. \ref{fig:results_all_ky_0} and \ref{fig:results_all_ky_1}
for the transverse wave vectors of $k_{y}=0$ and $k_{y}=1$, respectively.
For $k_{y}=0$ all the methods agree and only one curve is shown.
For $k_{y}=1$, and ``possibly'' all non-zero $k_{y}$, each method
yields a different result. It is easy to see in Fig. \ref{fig:results_all_ky_1}
that corrected Sadasivam's method and the method of decomposing the
Bloch matrices perform badly, as they yield transmissions that are
sometimes negative and sometimes above the maximum of unit transmission.
Huang's method, on the other hand, always gives transmission functions
that are positive and less than unity. This seems in the first place
that this method is correct. When inspected closely, however, one
can see that it gives non-vanishing transmission even when the dispersion
curve clearly shows that either of the corresponding modes are evanescent.
In light of these remarks, all three mentioned are deemed incorrect
when based solely on the numerical results. Our method which consists
of starting the decomposition process from the escape rates is the
only method that yield correct results when compared to the previously
published results. 

\subsection{Discussion}

It is remarkable that all the methods agree when the transverse momentum
$k_{y}=0$. This can be explained by: (a) the u's are orthogonal,
and (b) the system is effectively one-dimensional. In the following,
therefore, we limit our discussion to the non-vanishing transverse
momentum. This case is represented by the value $k_{y}=1$ considered
in the current numerical application.

We have seen from the results of the numerical calculations of the
mode-resolved transmission functions obtained by Huang \textit{et
al.}\cite{Huang11} that they are in disagreement from the ``correct''
results obtained using our method and the earlier correct ones. We
saw that the total transmission is always correct and the mode-resolved
functions are also well-behaved, meaning they are always positive
and less than or equal unity. These characteristics by themselves
do not allow to draw any conclusion as the correctness of the method.
The unmistakable flaw of the method, however, is shown when the mode-resolved
functions are non-zero (do not vanish) even when either mode is evanescent.
This feature does not even call to comparison to other methods, the
result being utterly unphysical. Moreover, Huang \textit{et al.}'s
method(Method I)\cite{Huang11} yields results that are incorrect
when compared to our method, even when both modes are propagating.
It is clear that the problem lays in the way the decomposition is
carried out within this method. Indeed, and as mentioned above there
is no obvious connection between eigenvalues $\lambda_{i}$ obtained
the $i^{\mathrm{th}}$ mode in the lead, and as such the resulting
$t_{i,j}$ cannot represent genuine mode-resolved transmission functions.

If we now use this corrected mode-resolved surface GF (Method II),
the results are even worse. We obtain for instance negative transmission
functions and functions that are greater than unity, even though the
total transmission is still correct. The last observation is the difference
with the results obtained using the original formulation derived in
\cite{Sadasivam17}. This is due to the total surface GF being identical
to the sum of the mode-resolved ones in (Method II); a condition missing
in the original formulation of the method. 

The apparently better method (Method III) still yields results that
in total disagreement with the methods of reference. This method is
also plagued with similar incorrect and physically nonsensical results
as those obtained from Sadasivam's method. Again, this tells us that
starting from mode-resolved surface GFs is not the right way to proceed.
From the last observations we note that even when carried out correctly
the resolution of surface GF into individual GFs yields incorrect
results. We conclude, therefore, that starting from resolved GFs is
not the right way to compute mode-resolved transmission functions

Our method (Method IV) when compared to the earlier method in \cite{Boumrar20}
yields identical results. We assert that the only way to obtain physically
correct and meaningful mode-resolved transmission functions is to
start from mode-resolved escape rates $\Gamma_{L/R}$. These decomposition
starts from group velocities, rather than from surface GFs. Using
Caroli's formula
\[
t_{\alpha,\beta}=\mathrm{Tr}\left[\Gamma_{L,\alpha}G_{c}\Gamma_{R,\beta}G_{c}^{\dagger}\right],
\]
it is clear, for instance, that if either mode is evanescent, the
resulting transmission vanishes. That this is the case is due to the
escape rates being computed directly from the group velocities. This,
at least, eliminates the possibility of obtaining non-zero transmission
when either mode is evanescent; a problem encountered in both Method
I and Method II.

\section{Conclusion}

In this work we have discussed several methods of calculating the
polarization-resolved transmittances in the phonon transport problem.
Each method carries out the decomposing from a different starting
point. The final step in all of them is obtaining mode-resolved escape
rates which are then plugged into the Caroli formula to yield mode-resolved
transmittances. While some methods are easily shown to be incorrect
by analytical means, most seem be to be inherently correct. That is
until confronted to the genuinely correct methods of decomposition
through numerical calculations. Even then, and for a set of given
parameters such $k_{y}=0$, the results may come out correct. This
a clear indication that when the tests are also carried out only for
a one-dimensional system, the result may give a misleading impression
of correctness. It is therefore also fortunate that one can draw unmistakable
conclusions from the value of each method by simulating such a simple
model as a square lattice. Our aim in choosing this model as mentioned
earlier is to make our calculations easy to reproduce, while making
our conclusions stand on a firm grounding. It is indeed unfortunate
that when a novel method is introduced more often than not the numerical
application is carried out on complicated systems with unpublished
parameters, rendering reproduction of the results a daunting task
to other workers not involved in the work. To conclude, we state that
of all the ways of decomposing the transmittance functions, the only
correct one is to start from decomposed escape rates which are written
directly in terms of ``generalized'' group velocities. 

\bibliographystyle{plain}
\bibliography{mainbibfile}

\end{document}